\documentclass[longauth]{aa}
\usepackage{graphicx}
\usepackage{txfonts}
\usepackage{xcolor}
\usepackage{hyperref}
\hypersetup{colorlinks, linkcolor=blue, citecolor=blue, urlcolor=blue}
\usepackage{amsmath}
\usepackage{ulem}
\usepackage{soul}
\usepackage{enumitem}

\newcommand{\pivec}{\mbox{\boldmath $\pi$}}
\newcommand{\muvec}{\mbox{\boldmath $\mu$}}

\newcommand{\te}{t_{\rm E}}
\newcommand{\thetae}{\theta_{\rm E}}

\newcommand{\pie}{\pi_{\rm E}}
\newcommand{\pien}{\pi_{{\rm E},N}}
\newcommand{\piee}{\pi_{{\rm E},E}}
\newcommand{\dl}{D_{\rm L}}
\newcommand{\ds}{D_{\rm S}}

\definecolor{brown}{rgb}{0.59, 0.29, 0.0}
\definecolor{darkgreen}{rgb}{0.0, 0.42, 0.24}
\definecolor{darkblue}{rgb}{0.0, 0.25, 0.42}
\definecolor{blue}{rgb}{0.0,0.0,1.0}
\definecolor{green}{rgb}{0.0,1.0,0.0}

\begin{document}

\title{OGLE-2018-BLG-0971, MOA-2023-BLG-065, and OGLE-2023-BLG-0136: Microlensing 
events with prominent orbital effects} \titlerunning{Microlensing events with prominent 
orbital effects}

\author{
     Cheongho~Han\inst{\ref{inst1}} 
\and Andrzej~Udalski\inst{\ref{inst2}} 
\and Ian~A.~Bond\inst{\ref{inst3}}
\and Chung-Uk~Lee\inst{\ref{inst4}} 
\and Andrew~Gould\inst{\ref{inst5},\ref{inst6}}      
\\
(Leading authors)
\\
     Michael~D.~Albrow\inst{\ref{inst7}}   
\and Sun-Ju~Chung\inst{\ref{inst4}}      
\and Kyu-Ha~Hwang\inst{\ref{inst4}} 
\and Youn~Kil~Jung\inst{\ref{inst4}} 
\and Hyoun-Woo Kim\inst{\ref{inst4}}
\and Yoon-Hyun~Ryu\inst{\ref{inst4}} 
\and Yossi~Shvartzvald\inst{\ref{inst8}}   
\and In-Gu~Shin\inst{\ref{inst9}} 
\and Jennifer~C.~Yee\inst{\ref{inst9}}   
\and Hongjing~Yang\inst{\ref{inst10}}     
\and Weicheng~Zang\inst{\ref{inst9},\ref{inst10}}     
\and Sang-Mok~Cha\inst{\ref{inst4},\ref{inst11}} 
\and Doeon~Kim\inst{\ref{inst1}}
\and Dong-Jin~Kim\inst{\ref{inst4}} 
\and Seung-Lee~Kim\inst{\ref{inst4}} 
\and Dong-Joo~Lee\inst{\ref{inst4}} 
\and Yongseok~Lee\inst{\ref{inst4},\ref{inst11}} 
\and Byeong-Gon~Park\inst{\ref{inst4}} 
\and Richard~W.~Pogge\inst{\ref{inst6}}
\\
(The KMTNet Collaboration)
\\
     Przemek~Mr{\'o}z\inst{\ref{inst2}} 
\and Micha{\l}~K.~Szyma{\'n}ski\inst{\ref{inst2}}
\and Jan~Skowron\inst{\ref{inst2}}
\and Rados{\l}aw~Poleski\inst{\ref{inst2}} 
\and Igor~Soszy{\'n}ski\inst{\ref{inst2}}
\and Pawe{\l}~Pietrukowicz\inst{\ref{inst2}}
\and Szymon~Koz{\l}owski\inst{\ref{inst2}} 
\and Krzysztof~A.~Rybicki\inst{\ref{inst2},\ref{inst8}}
\and Patryk~Iwanek\inst{\ref{inst2}}
\and Krzysztof~Ulaczyk\inst{\ref{inst12}}
\and Marcin~Wrona\inst{\ref{inst2}}
\and Mariusz~Gromadzki\inst{\ref{inst2}}          
\and Mateusz~J.~Mr{\'o}z\inst{\ref{inst2}} 
\\
(The OGLE Collaboration)
\\
     Fumio~Abe\inst{\ref{inst13}}
\and Richard~Barry\inst{\ref{inst14}}
\and David~P.~Bennett\inst{\ref{inst14},\ref{inst15}}
\and Aparna~Bhattacharya\inst{\ref{inst13},\ref{inst14}}
\and Hirosame~Fujii\inst{\ref{inst13}}
\and Akihiko~Fukui\inst{\ref{inst16},}\inst{\ref{inst17}}
\and Ryusei~Hamada\inst{\ref{inst18}}
\and Yuki~Hirao\inst{\ref{inst18}}
\and Stela~Ishitani Silva\inst{\ref{inst15},\ref{inst19}}
\and Yoshitaka~Itow\inst{\ref{inst13}}
\and Rintaro~Kirikawa\inst{\ref{inst18}}
\and Naoki~Koshimoto\inst{\ref{inst20}}
\and Yutaka~Matsubara\inst{\ref{inst13}}
\and Shota~Miyazaki\inst{\ref{inst18}}
\and Yasushi~Muraki\inst{\ref{inst13}}
\and Greg~Olmschenk\inst{\ref{inst14}}
\and Cl{\'e}ment~Ranc\inst{\ref{inst21}}
\and Nicholas~J.~Rattenbury\inst{\ref{inst22}}
\and Yuki~Satoh\inst{\ref{inst18}}
\and Takahiro~Sumi\inst{\ref{inst18}}
\and Daisuke~Suzuki\inst{\ref{inst18}}
\and Mio~Tomoyoshi\inst{\ref{inst18}}
\and Paul~J.~Tristram\inst{\ref{inst23}}
\and Aikaterini~Vandorou\inst{\ref{inst14},\ref{inst15}}
\and Hibiki~Yama\inst{\ref{inst18}}
\and Kansuke~Yamashita\inst{\ref{inst18}}
\\
(The MOA Collaboration)
}

\institute{
      Department of Physics, Chungbuk National University, Cheongju 28644, Republic of Korea\label{inst1}                                                          
\and  Astronomical Observatory, University of Warsaw, Al.~Ujazdowskie 4, 00-478 Warszawa, Poland\label{inst2}                                                      
\and  Institute of Natural and Mathematical Science, Massey University, Auckland 0745, New Zealand\label{inst3}                                                    
\and  Korea Astronomy and Space Science Institute, Daejon 34055, Republic of Korea\label{inst4}                                                                    
\and  Max Planck Institute for Astronomy, K\"onigstuhl 17, D-69117 Heidelberg, Germany\label{inst5}                                                                
\and  Department of Astronomy, The Ohio State University, 140 W. 18th Ave., Columbus, OH 43210, USA\label{inst6}                                                   
\and  University of Canterbury, Department of Physics and Astronomy, Private Bag 4800, Christchurch 8020, New Zealand\label{inst7}                                 
\and  Department of Particle Physics and Astrophysics, Weizmann Institute of Science, Rehovot 76100, Israel\label{inst8}                                           
\and  Center for Astrophysics $|$ Harvard \& Smithsonian 60 Garden St., Cambridge, MA 02138, USA\label{inst9}                                                      
\and  Department of Astronomy and Tsinghua Centre for Astrophysics, Tsinghua University, Beijing 100084, China\label{inst10}                                       
\and  School of Space Research, Kyung Hee University, Yongin, Kyeonggi 17104, Republic of Korea\label{inst11}                                                      
\and  Department of Physics, University of Warwick, Gibbet Hill Road, Coventry, CV4 7AL, UK\label{inst12}                                                          
\and  Institute for Space-Earth Environmental Research, Nagoya University, Nagoya 464-8601, Japan\label{inst13}                                                    
\and  Code 667, NASA Goddard Space Flight Center, Greenbelt, MD 20771, USA\label{inst14}                                                                           
\and  Department of Astronomy, University of Maryland, College Park, MD 20742, USA\label{inst15}                                                                   
\and  Department of Earth and Planetary Science, Graduate School of Science, The University of Tokyo, 7-3-1 Hongo, Bunkyo-ku, Tokyo 113-0033, Japan\label{inst16}  
\and  Instituto de Astrof{\'i}sica de Canarias, V{\'i}a L{\'a}ctea s/n, E-38205 La Laguna, Tenerife, Spain\label{inst17}                                           
\and  Department of Earth and Space Science, Graduate School of Science, Osaka University, Toyonaka, Osaka 560-0043, Japan\label{inst18}                           
\and  Department of Physics, The Catholic University of America, Washington, DC 20064, USA\label{inst19}                                                           
\and  Department of Astronomy, Graduate School of Science, The University of Tokyo, 7-3-1 Hongo, Bunkyo-ku, Tokyo 113-0033, Japan\label{inst20}                    
\and  Sorbonne Universit\'e, CNRS, UMR 7095, Institut d'Astrophysique de Paris, 98 bis bd Arago, 75014 Paris, France\label{inst21}                                 
\and  Department of Physics, University of Auckland, Private Bag 92019, Auckland, New Zealand\label{inst22}                                                        
\and  University of Canterbury Mt.~John Observatory, P.O. Box 56, Lake Tekapo 8770, New Zealand\label{inst23}                                                      
}                                                                                                                                                       
\date{Received ; accepted}

\abstract
{}
{
We undertake a project to reexamine microlensing data gathered from high-cadence surveys. The
aim of the project is to reinvestigate lensing events with light curves exhibiting intricate 
anomaly features associated with caustics, yet lacking prior proposed models to explain these 
features. 
}
{
Through detailed reanalyses considering higher-order effects, we identify that accounting for 
orbital motions of lenses is vital in accurately explaining the anomaly features observed in the
light curves of the lensing events OGLE-2018-BLG-0971, MOA-2023-BLG-065, and OGLE-2023-BLG-0136. 
}
{
We estimate  the masses and distances to the lenses by conducting Bayesian analyses using the
lensing parameters of the newly found lensing solutions.  From these analyses, we identify that 
the lenses of the events OGLE-2018-BLG-0971 and MOA-2023-BLG-065 are binaries composed of M 
dwarfs, while the lens of OGLE-2023-BLG-0136 is likely to be a binary composed of an early 
K-dwarf primary and a late M-dwarf companion.  For all lensing events, the probability of the 
lens residing in the bulge is considerably higher than that of it being located in the disk.  
}
{}


\maketitle

\begin{table*}[t]
\caption{Coordinates of events.\label{table:one}}
\begin{tabular}{lllllll}
\hline\hline
\multicolumn{1}{c}{Event}                    &
\multicolumn{1}{c}{(RA, Dec)$_{\rm J2000}$}  &
\multicolumn{1}{c}{$(l, b)$}                 &
\multicolumn{1}{c}{Other ID references}     \\
\hline
 OGLE-2018-BLG-0971  &  (17:59:01.63, -28:13:42.20)   &  $(2^\circ\hskip-2pt .1082, -2^\circ\hskip-2pt .1727)$   &  MOA-2018-BLG-173, KMT-2018-BLG-2336  \\ 
 MOA-2023-BLG-065    &  (18:00:35.68, -29:13:23.95)   &  $(1^\circ\hskip-2pt .4140, -2^\circ\hskip-2pt .9645)$   &  KMT-2023-BLG-2430  \\ 
 OGLE-2023-BLG-0136  &  (18:09:16.97, -30:36:47.00)   &  $(1^\circ\hskip-2pt .1132, -5^\circ\hskip-2pt .2851)$   &  KMT-2023-BLG-2849  \\ 
\hline
\end{tabular}
\end{table*}

\section{Introduction}\label{sec:one}

In general, light curves of microlensing events are modeled by assuming a rectilinear relative
motion between the lens and the source. However, deviations from this assumption arise due to
accelerations affecting the motion of either the observer, lens, or source. For instance, an 
observer experiences acceleration due to the Earth's orbital motion around the Sun, known as 
microlens-parallax effects \citep{Gould1992, Gould2000}. Additionally, if a source of a lensing 
event is part of a binary system in which two stars orbit a common barycenter, the source's 
motion also undergoes acceleration \citep{Han1997, Rahvar2009}. Similarly, the lens's orbital 
motion induces acceleration, causing deviations from a rectilinear relative motion between the 
lens and source, known as lens-orbital effects.

There have been instances of lensing events in which accounting for lens orbital motions was 
crucial for accurately interpreting observed lensing light curves. MACHO 97-BLG-41 
\citep{Alcock2000} notably marked the first binary-lens single-source (2L1S) system displaying 
significant deviations from the assumption of a static binary configuration. Initially, these 
deviations were attributed to the presence of a third body of the lens, specifically a 
circumbinary planet \citep{Bennett1999}.  However, \citet{Albrow2000}, based on  the analysis 
using an independent data set, later proposed a solution involving an orbiting binary lens. The 
controversy was definitively resolved by \citet{Jung2013}, who, through a direct comparison of 
the two models using a combined data set, found that the orbiting binary-lens interpretation is 
preferred over the circumbinary planet model.

OGLE-2006-BLG-109 was the second instance in which lens orbital effects played a crucial role 
in providing accurate explanations for the lensing light curve \citep{Gaudi2008, Bennett2010}.
The light curve of this event exhibited a complex anomaly pattern, comprising multiple 
distinctive features. A comprehensive understanding of these anomalies was achieved only after 
accounting for combined higher-order effects resulting from the parallactic motion of Earth and 
the orbital motion of the lens. From the analysis considering these higher-order effects, the 
lens was proven to be the first double planet system discovered with the gravitational 
microlensing method.

OGLE-2005-BLG-018 marked the third instance in which the significance of lens orbital effects 
was established. The light curve of the event displayed multiple anomaly features, comprising 
two neighboring strong anomalies and a comparatively weak anomaly positioned apart from the 
stronger ones. Although a model based on a static binary lens configuration could approximately
explain the two strong anomalies, accurately describing the separate weak anomaly was
challenging. As a result, this event had not been previously addressed until \citet{Shin2011}
revisited it and demonstrated that accounting for the lens orbital motion was essential for
accurately describing all the anomaly features in the lensing light curve.

Prompted by the work of \citet{Shin2011}, \citet{Park2013} revisited microlensing data available
until then and conducted thorough analyses of binary-lens events. They focused on cases in which
static binary models fell short in accurately describing observed light curves. Through these 
analyses, they revealed that the substantial residuals of the light curves of two lensing events
OGLE-2006-BLG-277 and OGLE-2012-BLG-0031 from static 2L1S models were predominantly
attributed to the influence of lens orbital motion.

OGLE-2009-BLG-020 was identified as a binary-lens event where orbital motion was initially predicted 
through light curve analysis \citep{Skowron2011}. Subsequent 3.5 years of radial velocity monitoring 
confirmed an orbit consistent with the predictions derived from the microlensing light curve analysis 
\citep{Yee2016}.

In the case OGLE-2013-BLG-0723, the lensing light curve was initially interpreted by a 
triple-lens model, in which the lens consisted of a Venus-mass planet and binary brown-dwarf host 
\citep{Udalski2015a}.  Later, \citet{Han2016a} reexamined the event, incorporating lens-orbital 
effects, and proposed a revised interpretation involving two-body lens instead of the previously 
suggested three-body lens solution. This updated model provided a notably better fit to the 
observed light curve.

Gaia16aye, detected toward the northern Galactic disk field, was a binary microlensing event and
stood among the earliest instances detected from the alerts issued by the Gaia space mission. The
brightening of the source induced by lensing endured for more than two years, and thus it was
crucial to include the orbital motion of the lens for the precise description of the lensing light
curve \citep{Wyrzykowski2020}.

The light curve of the lensing event KMT-2021-BLG-0322 exhibited multiple sets of caustic-crossing
features. The overall features of the light curve were approximately described by a 2L1S model, but
the model left substantial residuals. From the reanalysis, \citet{Han2021} found that the residuals
could be explained either by considering a non-rectilinear lens-source motion caused by the
combination of microlens-parallax and lens-orbital effects or by adding an additional low-mass
companion to the binary lens, and hence three lens components (3L1S system). The degeneracy
between the higher-order 2L1S model and the 3L1S model was very severe, making it difficult to
single out a correct solution based on the photometric data. This degeneracy was known before for
two previous events (MACHO-97-BLG-41 and OGLE-2013-BLG-0723), which led to the false detections 
of planets in binary systems, and thus the identification of the degeneracy for the event 
illustrated that this degeneracy could be common.

In this study, we present comprehensive analyses on three 2L1S lensing events: 
OGLE-2018-BLG-0971, MOA-2023-BLG-065, and OGLE-2023-BLG-0136. These events bear resemblances, 
displaying anomalies in their light curves with intricate and complex features that prove 
challenging to interpret using static binary-lens models. We demonstrate the importance of 
considering orbital motions of lenses in precisely describing the observed anomaly features in 
the lensing light curves.

\section{Event selections and data}\label{sec:two}

We undertook a project involving the reexamination of microlensing data obtained from three
ongoing high-cadence microlensing surveys: the Korea Microlensing Telescope Network 
\cite[KMTNet:][]{Kim2016}, the Optical Gravitational Lensing Experiment \citep[OGLE:][]{Udalski2015a}, 
and the Microlensing Observations in Astrophysics survey \citep[MOA:][]{Bond2001}. In this 
project, we directed our attention to lensing events displaying intricate anomaly features 
related to caustics, yet lacking prior proposed models to explain these features.  We commenced 
our investigation by analyzing the KMTNet data spanning from the 2016 season to the 2023 season, 
specifically focusing on identifying anomalous lensing events characterized by conspicuous 
caustic-crossing features. Subsequently, we examined the presence of lensing models corresponding 
to these events, filtering out candidate events for which either no models were proposed or 
presented models failed to accurately delineate the anomalous features. Finally, in the 
concluding step, we verified the availability of additional data obtained from the OGLE and 
MOA surveys for the candidate events.  Through this process, we identified three lensing 
events for which the consideration of orbital motions of lenses played crucial role in 
accurately describing the observed anomaly features in the lensing light curves, including 
OGLE-2018-BLG-0971, MOA-2023-BLG-065, and OGLE-2023-BLG-0136. In Table~\ref{table:one}, we 
present the equatorial and Galactic coordinates of the events. All these events were captured 
and observed by multiple surveys. We provide ID references assigned by the respective surveys, 
using the ID references from the initial discovery surveys in subsequent discussions.

Observations of the events by the individual surveys were conducted using the telescopes 
operated by the respective survey groups. The KMTNet group utilizes three identical telescopes, 
each featuring a 1.6-meter aperture and equipped with a camera capable of capturing a field 
spanning 4 square degrees. For continuous coverage of lensing events, the KMTNet telescopes 
are strategically distributed throughout three countries in the Southern Hemisphere: at the 
Siding Spring Observatory in Australia (KMTA), the Cerro Tololo Interamerican Observatory in 
Chile (KMTC), and the South African Astronomical Observatory in South Africa (KMTS). The MOA 
survey employs a telescope with a 1.8 m aperture located at the Mt. John Observatory in New 
Zealand.  The camera mounted on the MOA telescope has the capacity to capture a 2.2 square 
degree area of the sky in a single shot. The OGLE survey operates the 1.3-meter Warsaw telescope 
situated at the Las Campanas Observatory in Chile. The camera mounted on the OGLE telescope 
provides a field of view that spans 1.4 square degrees.  The primary observations conducted by 
the KMTNet and OGLE surveys were done in the $I$ band, whereas observations by the MOA survey 
were conducted in the custom MOA-$R$ band. In all surveys, a portion of images was acquired 
in the $V$ band for the color measurements of source stars.

The data of the events were processed using photometry pipelines that are customized to the
individual survey groups: KMTNet employed the \citet{Albrow2009} pipeline, OGLE utilized the
\citet{Udalski2003} pipeline, and MOA employed the \citet{Bond2001} pipeline.  For the use 
of the optimal data, the KMTNet data set was refined through a re-reduction process using the 
code developed by \citet{Yang2023}.  For each data set, error bars estimated from the photometry 
pipelines were recalibrated not only to ensure consistency of the errorbars with the scatter of 
data but also to set the $\chi^2$ value per degree of freedom (d.o.f.) for each data to unity. 
This normalization process was done in accordance with the procedure outlined by \citet{Yee2012}.

\section{Light curve modeling}\label{sec:three}

The light curves of all the analyzed events show anomalous features that are characteristic of 
caustics. Caustics arise when a lens system comprises multiple masses, and they represent the
source positions at which the magnification of a point source diverges to infinity.  Consequently,
the presence of caustic-related features in the light curves implies the involvement of lenses
composed of multiple masses in producing these events.

Taking the caustic-related features into account, our analysis begins by modeling the light 
curves within a static 2L1S framework. This framework operates under the assumption of rectilinear 
relative motion between the lens and the source. Within this static binary-lens model, a lensing 
light curve is defined by seven basic parameters. The first three of these parameters characterizes 
the source's approach to the lens, denoted as $(t_0, u_0, \te)$. These parameters represent the time 
of the closest lens-source approach, the separation at that instant (impact parameter), and the event 
time scale, respectively. Here $u_0$ is scaled to the angular Einstein radius $\thetae$, and $\te$ is 
defined as the duration for the source to traverse the Einstein radius. The additional set of three 
parameters $(s, q, \alpha)$ characterizes the binary lens configuration. These parameters represent 
the projected separation (scaled to $\thetae$), the mass ratio between the binary lens components 
($M_1$ and $M_2$), and the angle (source trajectory angle) formed between the lens-source proper 
motion vector $\muvec$ and the axis defined by $M_1$ and $M_2$. The last parameter $\rho$, which 
is defined as the ratio of the angular source radius $\theta_*$ to the Einstein radius, that is, 
$\rho=\theta_*/\thetae$, quantifies the deformation of lensing light curves during caustic crossings 
due to finite-source effects. In the 2L1S modeling, we begin by exploring the binary parameters $s$ 
and $q$ using a grid approach, employing multiple initial values of $\alpha$. Subsequently, we 
determine the remaining parameters through a downhill method based on the Markov Chain Monte Carlo 
(MCMC) technique. The lensing solutions are further refined by enabling variation in all parameters.

As shown in the following section, interpreting the light curves of the events examined in this
study solely through static 2L1S models poses a significant challenge. In such instances, we
undertake additional modeling to account for higher-order effects which induces deviations in the
relative lens-source motion from a rectilinear path. Our modeling approach explores the influences
of both lens orbital motion and microlens-parallax effects. To integrate these higher-order effects
into the modeling, additional parameters beyond the fundamental set need to be included. The
additional parameters introduced for modeling with microlens-parallax effects encompass 
$(\pien, \piee)$, representing the north and east components of the microlens-parallax vector, 
$\pivec_{\rm E}$, respectively.  The microlens-parallax vector is defined as
\begin{equation}
\pivec_{\rm E} = \left( {\pi_{\rm rel}\over \thetae} \right) \left( {\muvec \over \mu} \right),
\label{eq1}
\end{equation}
where $\pi_{\rm rel} = {\rm au}(1/\dl - 1/\ds)$ denotes the relative lens-source parallax, and 
$(\dl, \ds)$ represent the distances to the lens and source, respectively.  Under the first-order 
approximation of small changes in the positions of the lenses during lensing magnifications, the 
lens-orbital effects can be characterized by two parameters: $(ds/dt, d\alpha/dt)$.  These parameters 
represent the rates of change in the binary separation and the source trajectory angle, respectively.  
In the parallax+orbit modeling, we enforced a condition where the projected kinetic-to-potential 
energy ratio was required to remain below unity.  This condition ensured that the planet remained 
gravitationally bound to its host.  The energy ratio was computed from the higher-order lensing 
parameters by 
\begin{equation}
\left( {{\rm KE}\over {\rm PE}}\right)_\perp = 
{(a_\perp/{\rm au})^3 \over 8\pi^2(M/M_\odot)}
\left[
\left( {1\over s} {ds/dt\over {\rm yr}^{-1}}\right)^2+
\left( {d\alpha/dt \over {\rm yr}^{-1}}\right)^2
\right].
\label{eq2}
\end{equation}
Here $a_\perp$ denotes the physical separation between the binary lens components.  
Fully describing the Keplerian orbital motion necessitates the inclusion of additional 
parameters.\footnote{To access a comprehensive description of the orbital lensing 
parameters, refer to the summary provided in the Appendix of \citet{Skowron2011}.} However, 
determining these extra parameters poses a challenge due to the limitations of gravitational 
lensing, which is insensitive to the motion of the lens along the line of sight together with 
the partial light curve coverage spanning only  a minor fraction of the rotation period 
\citep{Albrow2000}.

Another higher-order effect that causes accelerations of the relative lens-source motion is 
the orbital motion of the source, known as the "xallarap effect" \citep{Griest1992, Han1997, 
Poindexter2005, Zhu2017, Satoh2023}. While the lens is confirmed to be binary, there is no 
prior justification to assume a binary source, although this possibility cannot be entirely 
dismissed. Therefore, we refrain from testing the xallarap effect as long as the observed 
anomalies can be explained by lens-orbital effects.

On occasion, light curves of 2L1S events affected by higher-order effects can imitate the patterns 
seen in 3L1S event light curves.  This confusion typically arises when a distinct anomaly feature 
in a lensing light curve is isolated from the primary ones, as seen in previous events such as 
MACHO-97-BLG-41, OGLE-2013-BLG-0723, and KMT-2021-BLG-0322.  For the analyzed events in this work, 
we find that such degeneracies do not exist.  In the following subsections, we offer detailed 
analyses of each individual event.

\begin{table*}[t]
\caption{Lensing parameters of OGLE-2018-BLG-0971\label{table:two}}
\begin{tabular}{lllllll}
\hline\hline
\multicolumn{1}{c}{Parameter}         &
\multicolumn{1}{c}{Static}            &
\multicolumn{1}{c}{Orbit}             &
\multicolumn{1}{c}{Orbit + Parallax}  \\
\hline
  $\chi^2$                      &  $13921.7               $   &    $9973.2                $    &  $9972.3               $    \\ 
  $t_0$ (${\rm HJD}^\prime$)    &  $58278.5148 \pm 0.0085 $   &    $58279.0241 \pm 0.0098 $    &  $58279.0245 \pm 0.0097$    \\ 
  $u_0$                         &  $0.2771 \pm 0.0006     $   &    $0.2956 \pm 0.0005     $    &  $0.2955 \pm 0.0008    $    \\ 
  $\te$ (days)                  &  $7.143 \pm 0.012       $   &    $7.126 \pm 0.009       $    &  $7.128 \pm 0.011      $    \\ 
  $s  $                         &  $0.9552 \pm 0.0006     $   &    $1.0070 \pm 0.0009     $    &  $1.0074 \pm 0.0010    $    \\ 
  $q  $                         &  $0.7161 \pm 0.0062     $   &    $0.8669 \pm 0.0064     $    &  $0.8653 \pm 0.0066    $    \\ 
  $\alpha$ (rad)                &  $2.1609 \pm 0.0024     $   &    $2.2964 \pm 0.0027     $    &  $2.2968 \pm 0.0025    $    \\ 
  $\rho$ ($10^{-2}$)            &  $1.078 \pm 0.017       $   &    $1.218 \pm 0.011       $    &  $1.221 \pm 0.011      $    \\ 
  $\pien$                       &   --                        &     --                         &  $0.07 \pm 0.27        $    \\   
  $\piee$                       &   --                        &     --                         &  $-0.016 \pm 0.091     $    \\
  $ds/dt$ (yr$^{-1}$)           &   --                        &    $-1.722 \pm 0.048      $    &  $-1.720 \pm 0.094     $    \\
  $d\alpha/dt$ (rad yr$^{-1}$)  &   --                        &    $0.931 \pm 0.026       $    &  $0.939 \pm 0.054      $    \\
\hline                                                                                                                 
\end{tabular}
\end{table*}

\begin{figure}[t]
\includegraphics[width=\columnwidth]{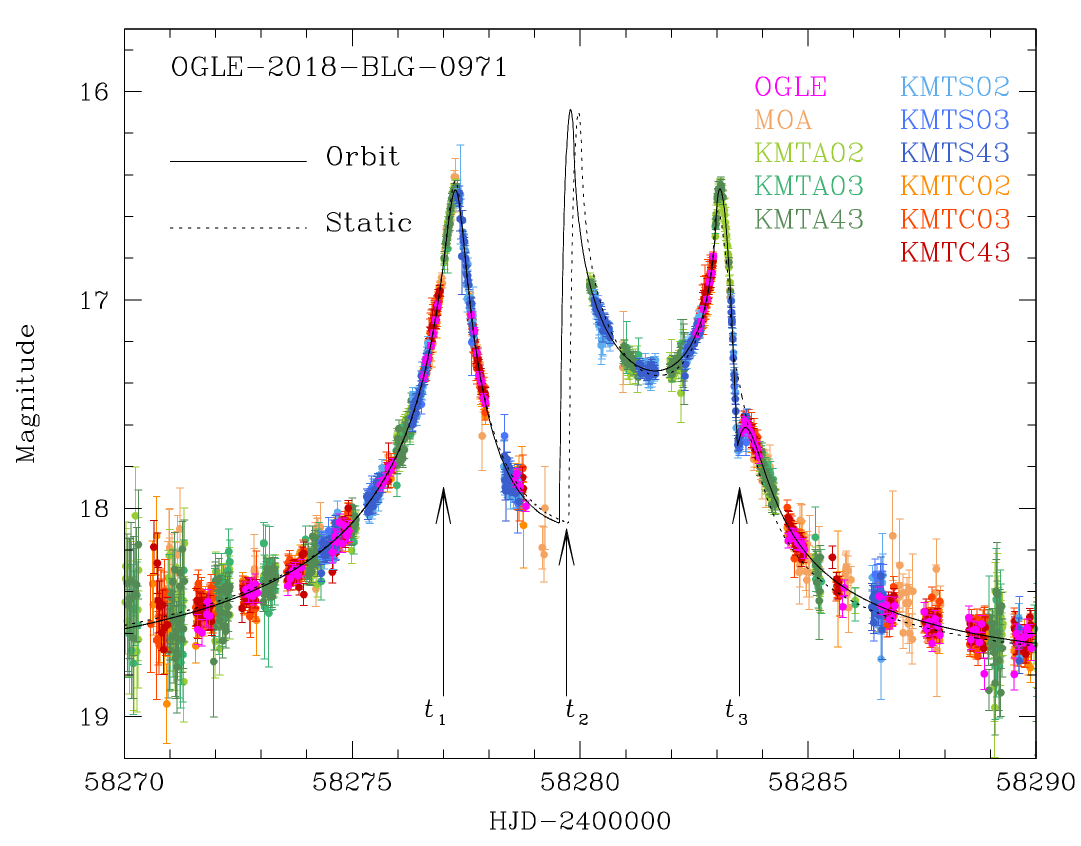}
\caption{
Light curve of the lensing event OGLE-2018-BLG-0971. The solid and dotted curves drawn over data 
points represent the model curves obtained from the 2L1S analyses with and without considering the 
lens orbital motion, respectively. The arrows with labels $t_1$, $t_2$, and $t_3$ denote the times 
of the major anomaly features.
}
\label{fig:one}
\end{figure}

\subsection{OGLE-2018-BLG-0971}\label{sec:three-one}

\begin{figure}[t]
\includegraphics[width=\columnwidth]{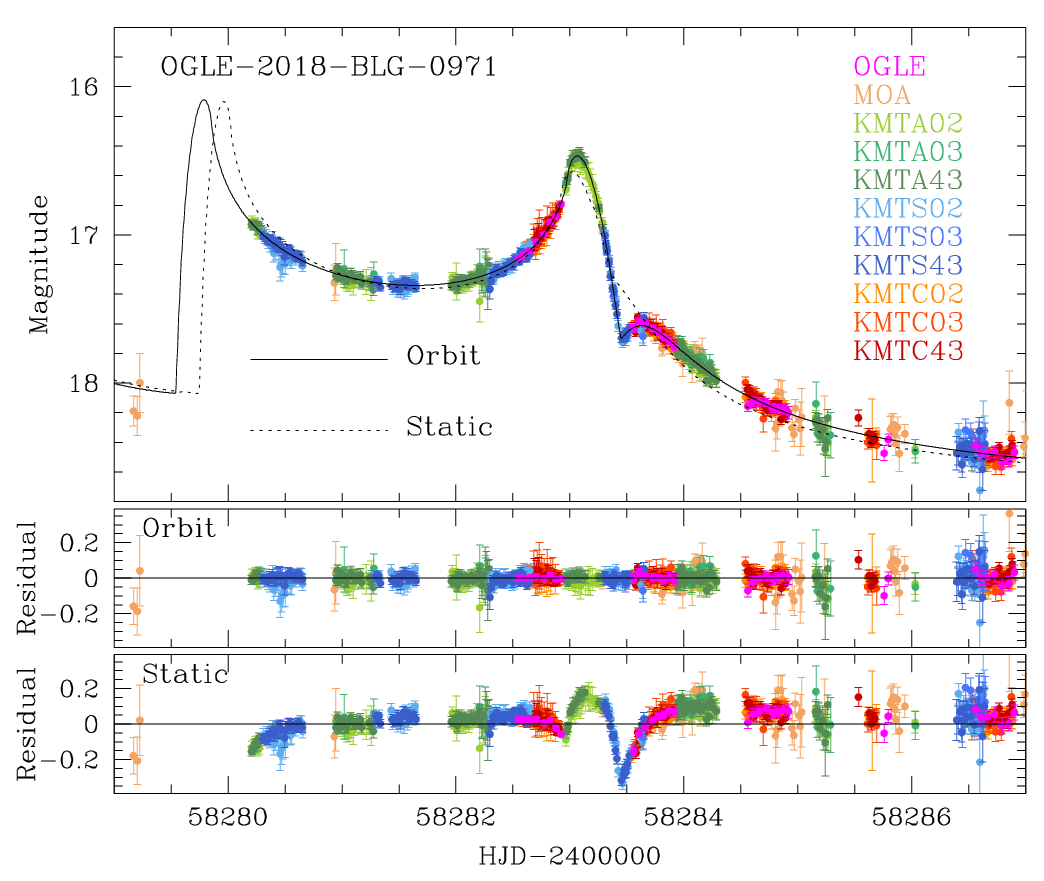}
\caption{
Enlarged view of the OGLE-2018-BLG-0971 light curve in the region around $t_2$ and $t_3$ 
marked in Fig.~\ref{fig:two}. 
}
\label{fig:two}
\end{figure}

The OGLE group first identified the lensing event OGLE-2018-BLG-0971 on 4 June, 2018.  Four 
days later, the MOA group verified the event, and the KMTNet group later retrieved it from a 
post-season data examination. The MOA and KMTNet groups designated the event as MOA-2018-BLG-173 
and KMT-2018-BLG-2336, respectively.  Figure~\ref{fig:one} shows the light curve of the event. 
The source of the event was situated within the overlapping region of the three KMTNet fields -- 
BLG02, BLG03, and BLG43. To differentiate between the individual data sets, we designate labels 
corresponding to the respective fields. The light curve manifests multiple anomaly features 
centered at ${\rm HJD}\equiv {\rm HJD}-2400000 \sim 58277$ ($t_1$), $\sim 58280$ ($t_2$), and 
$\sim 58283$ ($t_3$). The symmetric pattern with respect to $t_1$ suggests that the anomaly 
around this epoch likely stems from the source's approach to a caustic cusp, while the U-shaped
pattern spanning $t_2$ to $t_3$ indicates that these epochs correspond to the times of caustic 
entrance and exit. Apart from these anomaly features, there is an additional subtle anomaly 
feature appearing just after the caustic exit. Figure~\ref{fig:two} offers a detailed view of 
this specific region.

After modeling the light curve using a static 2L1S framework, we identified a solution that
broadly captures the features of the anomaly. In Table~\ref{table:two}, we list the lensing 
parameters of the static solution. However, it is found that the static model exhibits subtle 
residuals, particularly in the vicinity of $t_3$, as illustrated in the magnified view presented 
in Figure~\ref{fig:two}.

\begin{figure}[t]
\includegraphics[width=\columnwidth]{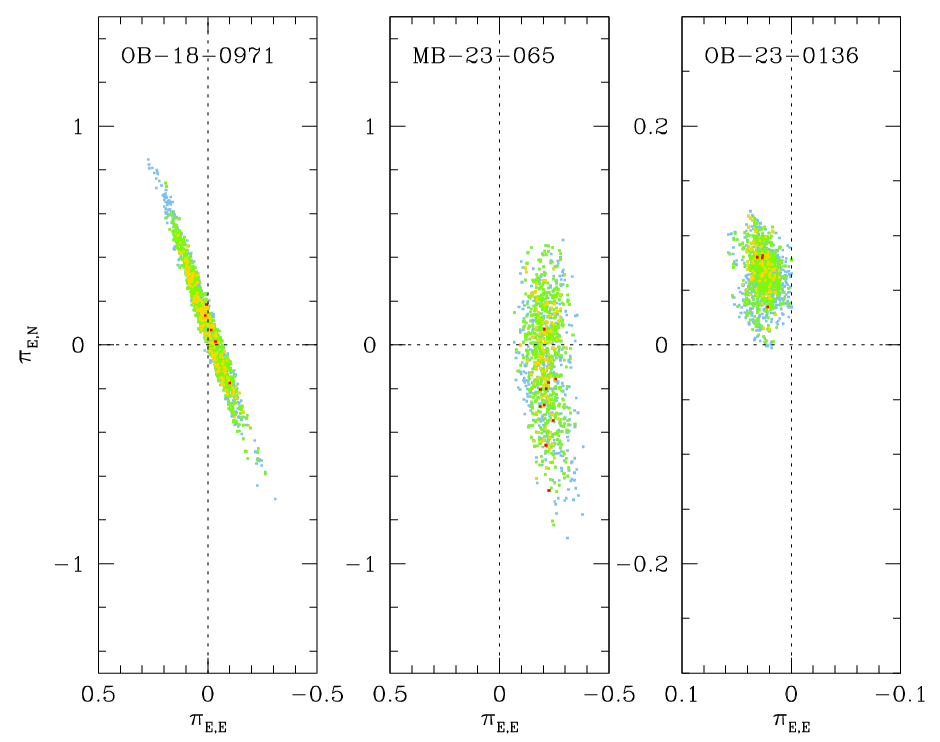}
\caption{
Scatter plots of points in the MCMC chain on the $\piee$--$\pien$ planes for the four events 
analyzed in this paper.  Points are color-coded to represent those with $<1\sigma$ (red), 
$<2\sigma$ (yellow), $<3\sigma$ (green), and $<4\sigma$ (cyan).  
}
\label{fig:three}
\end{figure}

\begin{figure}[t]
\includegraphics[width=\columnwidth]{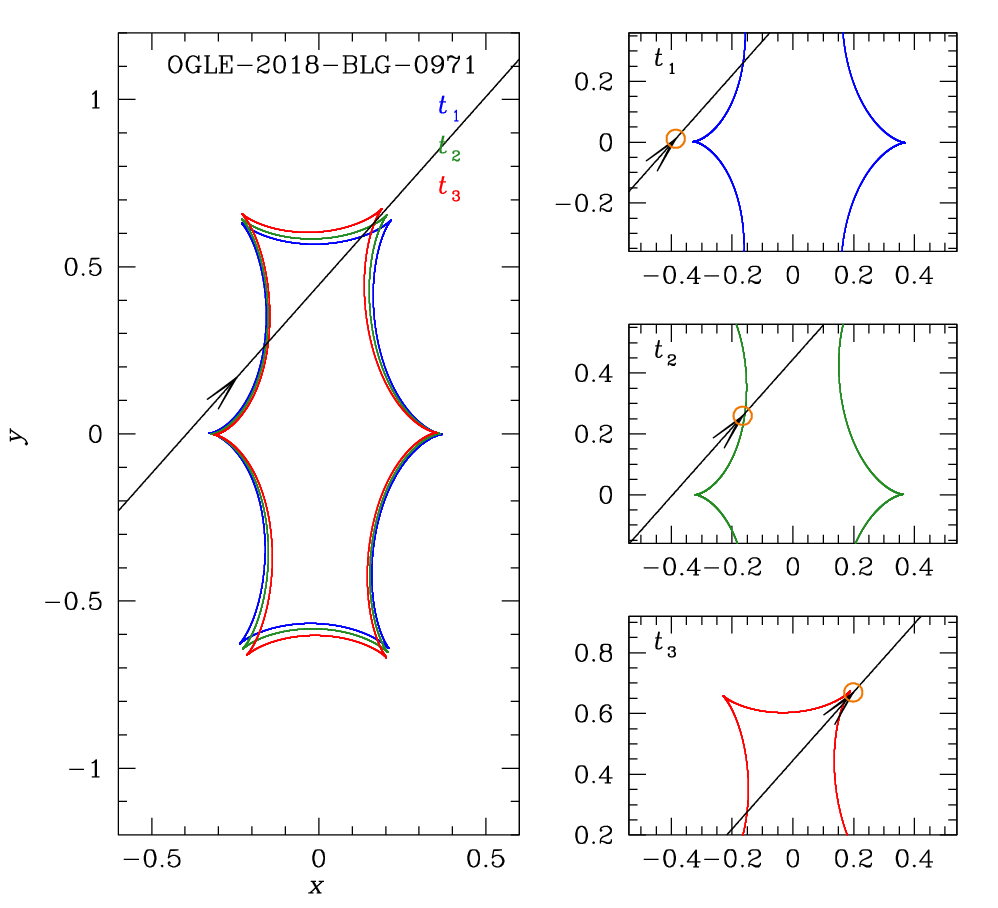}
\caption{
Lens system configuration for OGLE-2018-BLG-0971. 
The arrowed line represents 
the source trajectory, and the closed figures composed of concave curves represent caustics. 
The caustic shape and location evolve over time due to the orbital motion of the lens. The sets 
of caustics drawn in black, green, and red correspond to the three anomaly epochs $t_1$, $t_2$, 
and $t_3$ marked in Fig.~\ref{fig:one}. The right panels display still frames capturing the 
source's approach to the caustic at these three epochs. The orange circle on the source trajectory 
appearing in each right panel is sized to indicate the angular dimension of the source relative 
to the size of the caustic. 
}
\label{fig:four}
\end{figure}

In light of deviations from the static model, we checked the feasibility of explaining the deviation
with higher-order effects. The event duration, estimated at $\te \sim 7.1$ days from the static model, 
is short.  Consequently, we initially considered the lens-orbital effect in the modeling.  Through 
this approach, we derived a solution that accounts for all anomaly features. The model curve of this 
solution is shown as a solid line in Figure~\ref{fig:one}, offering a comprehensive view, and in 
Figure~\ref{fig:two}, providing an enlarged view around $t_3$.  The fit significantly improved with 
the inclusion of lens orbital motion, by $\Delta\chi^2 = 3949.4$ compared to the static model. 
Subsequent examination of microlens-parallax effects through additional modeling revealed a minimal 
fit improvement, by $\Delta\chi^2=0.9$, indicating the predominant influence of lens-orbital effects.  
In Figure~\ref{fig:three}, we plot the scatter plot of points in the MCMC chain on the $\piee$--$\pien$ 
plane.  In Table~\ref{table:two}, we provide the lensing parameters for both the orbit-only and 
orbit+parallax solutions. The parameters defining the binary lens are $(s, q) \sim (1.01, 0.87)$, 
indicating that the event was generated a binary system composed of roughly equal masses with a 
separation close to the Einstein radius of the lens system.  Although the time scale of the event, 
$\te\sim 7.1$~days, represents only a small fraction of the Earth's orbital period, we conducted 
separate modeling specifically to account for the microlens-parallax effect.  From this analysis, it 
was observed that not only did the fit perform worse compared to the orbit-only model by $\Delta\chi^2
=42.6$, but also the derived parallax parameters $(\piee, \pien)\sim (35.80, 32.50)$ appeared  absurdly 
large for a typical Galactic lensing event.

The configuration of the lens system for the lensing event OGLE-2018-BLG-0971 is shown in
Figure~\ref{fig:four}. This configuration reveals that the lens system forms a single set of resonant 
caustics featuring six cusps -- two along the binary axis and four positioned away from the axis. The
source initially approached the left on-axis cusp around $t_1$, entered the caustic near $t_2$, and 
exited the caustic at around $t_3$.  These caustic approach and crossings gave rise to anomalies at 
the corresponding epochs.  The primary deviation of the static 2L1S model, particularly around $t_3$, 
stems from its inability to account for the variation in the caustic caused by the orbital motion of 
the binary lens.

\begin{table*}[t]
\caption{Lensing parameters of MOA-2023-BLG-065\label{table:three}}
\begin{tabular}{lllllll}
\hline\hline
\multicolumn{1}{c}{Parameter}          &
\multicolumn{1}{c}{Static}             &
\multicolumn{1}{c}{Parallax}           &
\multicolumn{1}{c}{Orbit}              &
\multicolumn{1}{c}{Orbit + Parallax}   \\
\hline
  $\chi^2$                      &  $2399.0              $       &   $1876.6             $  &  $1882.7              $    &  $1872.6              $     \\ 
  $t_0$ (${\rm HJD}^\prime$)    &  $60030.308 \pm 0.024 $       &   $60030.140 \pm 0.045$  &  $60029.951 \pm 0.079 $    &  $60030.356 \pm 0.090 $     \\ 
  $u_0$                         &  $-0.3893 \pm 0.0053  $       &   $-0.426 \pm 0.009   $  &  $-0.4539 \pm 0.0079  $    &  $-0.4378 \pm 0.0083  $     \\ 
  $\te$ (days)                  &  $30.74 \pm 0.51      $       &   $37.27 \pm 0.59     $  &  $39.03 \pm 0.54      $    &  $37.81 \pm 0.75      $     \\ 
  $s  $                         &  $1.327 \pm 0.011     $       &   $1.452 \pm 0.008    $  &  $1.471 \pm 0.007     $    &  $1.467 \pm 0.010     $     \\ 
  $q  $                         &  $1.089 \pm 0.013     $       &   $0.877 \pm 0.041    $  &  $0.939 \pm 0.023     $    &  $0.887 \pm 0.022     $     \\ 
  $\alpha$ (rad)                &  $4.7757 \pm 0.0001   $       &   $4.7668 \pm 0.0015  $  &  $4.7705 \pm 0.0050   $    &  $4.7477 \pm 0.0062   $     \\ 
  $\rho$ ($10^{-3}$)            &  $2.018 \pm 0.039     $       &   $1.724 \pm 0.041    $  &  $1.763 \pm 0.033     $    &  $1.739 \pm 0.037     $     \\ 
  $\pien$                       &   --                          &   $0.069 \pm 0.119    $  &   --                       &  $-0.20 \pm 0.26      $     \\   
  $\piee$                       &   --                          &   $-0.157 \pm 0.045   $  &   --                       &  $-0.214 \pm 0.054    $     \\
  $ds/dt$ (yr$^{-1}$)           &   --                          &    --                    &  $0.015 \pm 0.071     $    &  $-0.296 \pm 0.082    $     \\
  $d\alpha/dt$ (rad yr$^{-1}$)  &   --                          &    --                    &  $-0.531 \pm 0.159    $    &  $0.217 \pm 0.506     $     \\
\hline                                                                                                                 
\end{tabular}
\end{table*}

\begin{figure}[t]
\includegraphics[width=\columnwidth]{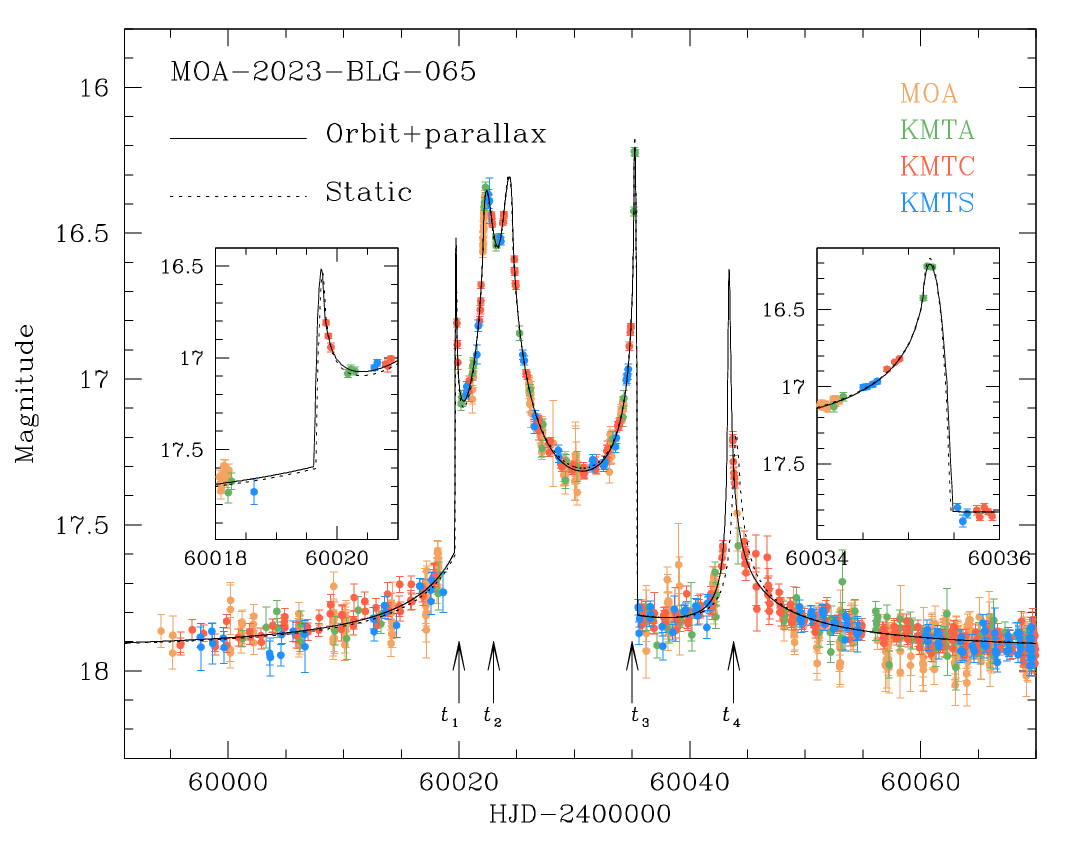}
\caption{
Lensing light curve of MOA-2023-BLG-065. The left and right insets show the zoom-in views of the 
regions around the caustic spikes at $t_1$ and $t_3$. 
}
\label{fig:five}
\end{figure}

\subsection{MOA-2023-BLG-065}\label{sec:three-two}

Figure~\ref{fig:five} shows the lensing light curve of the MOA-2023-BLG-065 event. The source flux 
magnification induced by lensing was initially identified on 17 March, 2023 (${\rm HJD}^\prime =60021$) 
through the survey conducted by the MOA group, subsequently confirmed by the KMTNet group.  The ID 
reference designated by the KMTNet survey is KMT-2023-BLG-2430.  The light curve displays a complex 
pattern comprising multiple anomaly features. Notably, two spike features at $t_1\sim 60020$ and 
$t_3\sim 60035$ appear to be a pair of caustic-crossing spikes.  Furthermore, the symmetry observed 
between the ascending and descending segments of the anomaly feature centered at $t_4\sim 60043$ 
implies that it originated from the source approach to a cusp of the caustic.  While the magnification 
between caustic spikes typically follows a U-shape pattern, the region between the caustic spikes at 
$t_1$ and $t_3$ deviates significantly from the U-shape, displaying a distinctive rise and fall in 
the region centered at $t_2\sim 60023$. This deviation is indicative of a source asymptotically 
approaching a fold of the caustic.

\begin{figure}[t]
\includegraphics[width=\columnwidth]{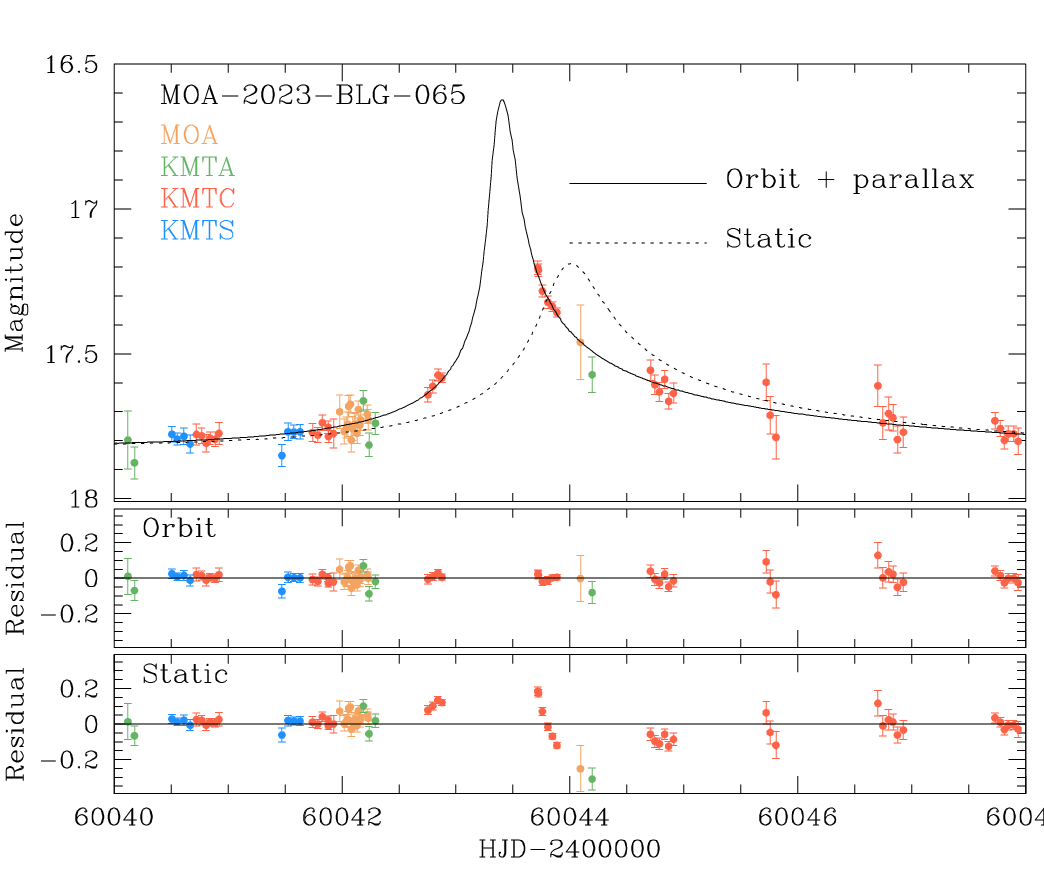}
\caption{
Enlarged view of the MOA-2023-BLG-065 light curve in the region around $t_4$ marked in Fig.~\ref{fig:five}. 
}
\label{fig:six}
\end{figure}

From the 2L1S analysis of the lensing light curve under a static binary frame, we found that the model 
falls short of precisely describing the data, even though it approximately outlines the anomaly features. 
In Figure~\ref{fig:five}, the static 2L1S model is represented by a dotted curve. This static model 
exhibits a notably inadequate fit especially in the region of the anomaly centered at $t_4$, as highlighted 
in the enlarged view presented in Figure~\ref{fig:six}. The complete lensing parameters for the static 
2L1S solution are detailed in Table~\ref{table:three}.

Although the static solution cannot adequately describe the anomaly feature at around $t_4$, it 
does exhibit an anomaly around the time of that anomaly.  This suggests the possibility that it 
can be described with a slight deformation of the source trajectory caused by higher-order effects.  
In light of this possibility, we conducted three additional sets of modeling: the first two models 
separately incorporated microlens-parallax and lens-orbital effects, while the third model encompassed 
both effects simultaneously.  In Table~\ref{table:three}, we present the lensing parameters for the 
three models.  Comparing the static and higher-order solution demonstrates a notable enhancement in 
fit, with a $\Delta\chi^2= 526.4$ compared to the static solution.  In Figure~\ref{fig:three}, we 
present the scatter plot of points in the MCMC chain on the $\piee$--$\pien$ plane.  It shows that 
the east component of the microlens-parallax vector, $\piee$, is constrained, although the uncertainty 
of the north component, $\pien$, is large.  Additionally, the normalized source radius $\rho = (1.739 
\pm 0.037) \times 10^{-3}$ was measured from the deformation of the light curve by finite source 
effects during the epochs around $t_2$ and $t_3$. The magnified views of these specific regions are 
shown in the two insets of Figure~\ref{fig:five}.

\begin{figure}[t]
\includegraphics[width=\columnwidth]{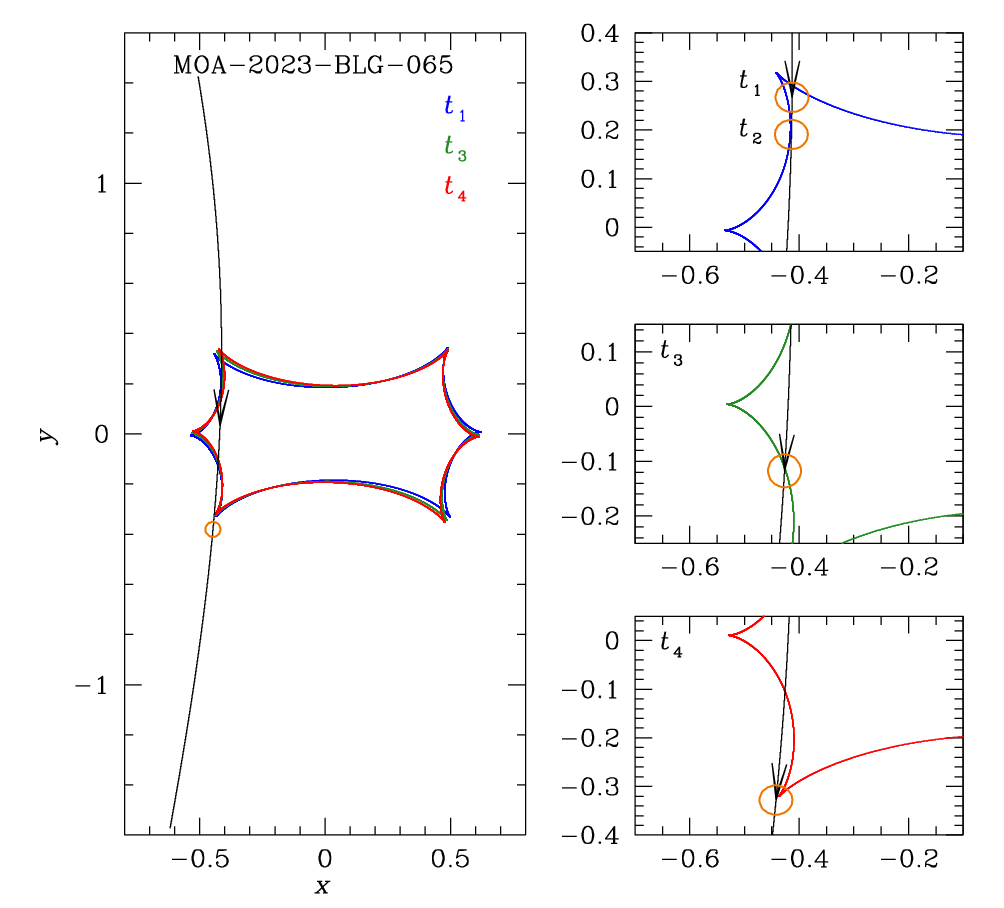}
\caption{
Lens system configuration for MOA-2023-BLG-065. 
}
\label{fig:seven}
\end{figure}

Figure~\ref{fig:seven} illustrates the lens-system configuration of MOA-2023-BLG-065. It shows that 
the binary lens, characterized by parameters $(s, q)\sim (1.5, 0.9)$, creates a resonant caustic, 
and the source traversed the left side of this caustic. Initially, the source passed the upper fold, 
creating the first caustic spike at $t_1$. Subsequently, it approached the upper left fold asymptotically, 
resulting in rising and falling features around $t_2$. The source then exited the caustic, generating 
the second caustic spike at $t_3$, and proceeded to pass the tip of the lower left cusp, causing the 
final anomaly feature at around $t_4$.  The distortion of the caustic due to the orbital motion of 
the lens led the light curve to deviate from the anticipated behavior according to the static model.  
This discrepancy was particularly noticeable in the part of the light curve during the final cusp 
approach, enabling the detection of the lens orbital effect.

\begin{table*}[t]
\caption{Lensing parameters of OGLE-2023-BLG-0136\label{table:four}}
\begin{tabular}{lllllll}
\hline\hline
\multicolumn{1}{c}{Parameter}         &
\multicolumn{1}{c}{Static}            &
\multicolumn{1}{c}{Parallax}          &
\multicolumn{1}{c}{Orbit}             &
\multicolumn{1}{c}{Orbit + Parallax}  \\
\hline
  $\chi^2$                      &  $3180.0              $   &    $2773.6                  $  &   $812.6               $    &   $809.4               $     \\ 
  $t_0$ (${\rm HJD}^\prime$)    &  $60030.726 \pm 0.093 $   &    $60029.065 \pm 0.132     $  &   $60030.276 \pm 0.107 $    &   $60030.08 \pm 0.11   $     \\ 
  $u_0$                         &  $0.2685 \pm 0.0007   $   &    $    0.2290 \pm 0.0008   $  &   $0.2808 \pm 0.0012   $    &   $0.2833 \pm 0.0017   $     \\ 
  $\te$ (days)                  &  $66.70 \pm 0.13      $   &    $   86.00 \pm 0.2999499  $  &   $61.25 \pm 0.13      $    &   $59.35 \pm 0.48      $     \\ 
  $s  $                         &  $0.66848 \pm 0.00042 $   &    $    0.68593 \pm 0.00034 $  &   $0.71297 \pm 0.00035 $    &   $0.71101 \pm 0.00085 $     \\ 
  $q  $                         &  $0.3117 \pm 0.0015   $   &    $    0.2607 \pm 0.0025   $  &   $0.2851 \pm 0.0027   $    &   $0.2978 \pm 0.0034   $     \\ 
  $\alpha$ (rad)                &  $1.4503 \pm 0.0009   $   &    $    1.3623 \pm 0.0019   $  &   $1.4268 \pm 0.0016   $    &   $1.42832\pm 0.0027   $     \\ 
  $\rho$ ($10^{-3}$)            &  $1.965 \pm 0.029     $   &    $    1.535 \pm 0.03      $  &   $1.992 \pm 0.029     $    &   $2.005 \pm 0.029     $     \\ 
  $\pien$                       &   --                      &    $    0.016 \pm 0.003     $  &    --                       &   $0.080  \pm 0.021    $     \\   
  $\piee$                       &   --                      &    $   -0.220 \pm 0.001     $  &    --                       &   $0.030 \pm 0.010     $     \\
  $ds/dt$ (yr$^{-1}$)           &   --                      &     --                         &   $-0.4403 \pm 0.0037  $    &   $-0.4410 \pm 0.0039  $     \\
  $d\alpha/dt$ (rad yr$^{-1}$)  &   --                      &     --                         &   $-0.237 \pm 0.007    $    &   $-0.408 \pm 0.050    $     \\
\hline                                                                                                                 
\end{tabular}
\end{table*}

\begin{figure}[t]
\includegraphics[width=\columnwidth]{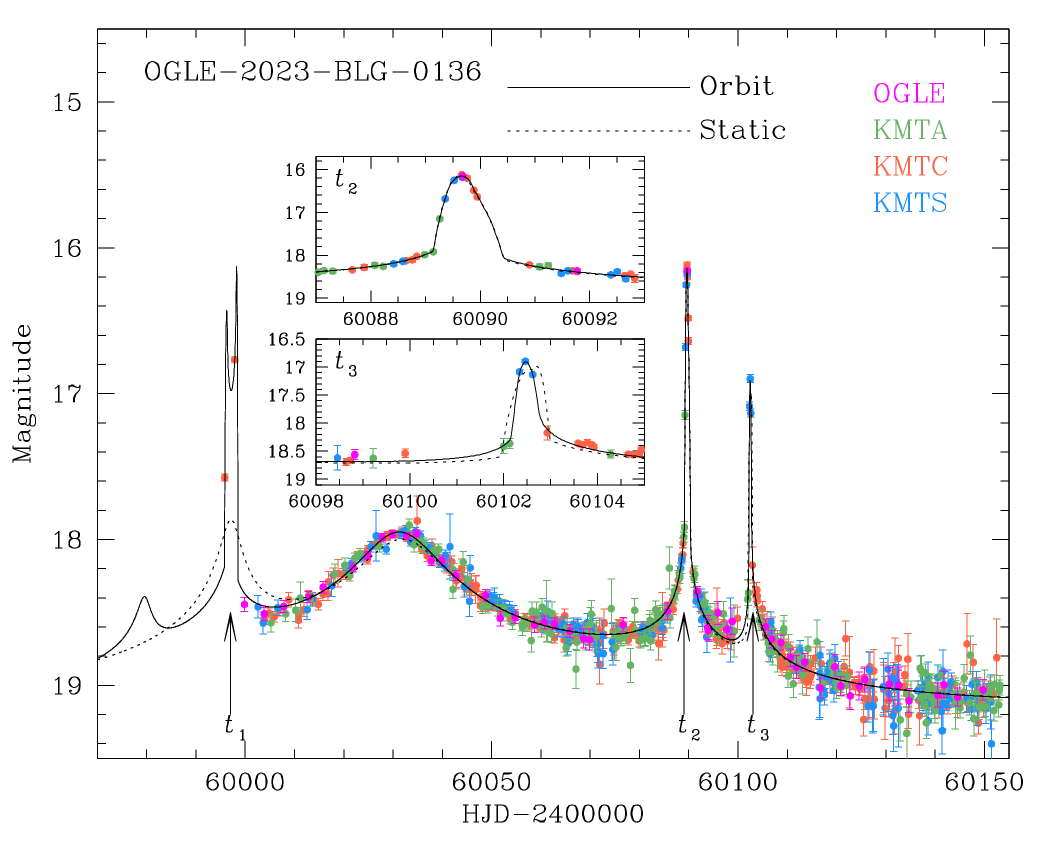}
\caption{
Lensing light curve of OGLE-2023-BLG-0136.  The insets provides closer look at the regions
surrounding $t_2$ and $t_3$. 
}
\label{fig:eight}
\end{figure}

\subsection{OGLE-2023-BLG-0136 }\label{sec:three-three}

The lensing event OGLE-2023-BLG-0136 was first found during its early phase by the OGLE group 
on 1 April, 2023 (${\rm HJD}^\prime =60036$).  Subsequently, the KMTNet survey validated the 
event, designating it as KMT-2023-BLG-2849.  Figure~\ref{fig:eight} illustrates the lensing 
light curve of the event. The curve exhibits a complex pattern of anomalies, comprised of 
multiple distinct features: a caustic-crossing feature around $t_1=59997$ and two additional 
features centered at around $t_2=60089$ and $t_3=60103$.  The first feature had limited coverage 
because it was observed solely by the KMTC telescope, while the other KMTNet telescopes and the 
OGLE telescope did not start observations of the 2023 season at the time of the anomaly. However, 
the subsequent two anomaly features were extensively observed by the combined data from both the 
OGLE and KMTNet surveys.

\begin{figure}[t]
\includegraphics[width=\columnwidth]{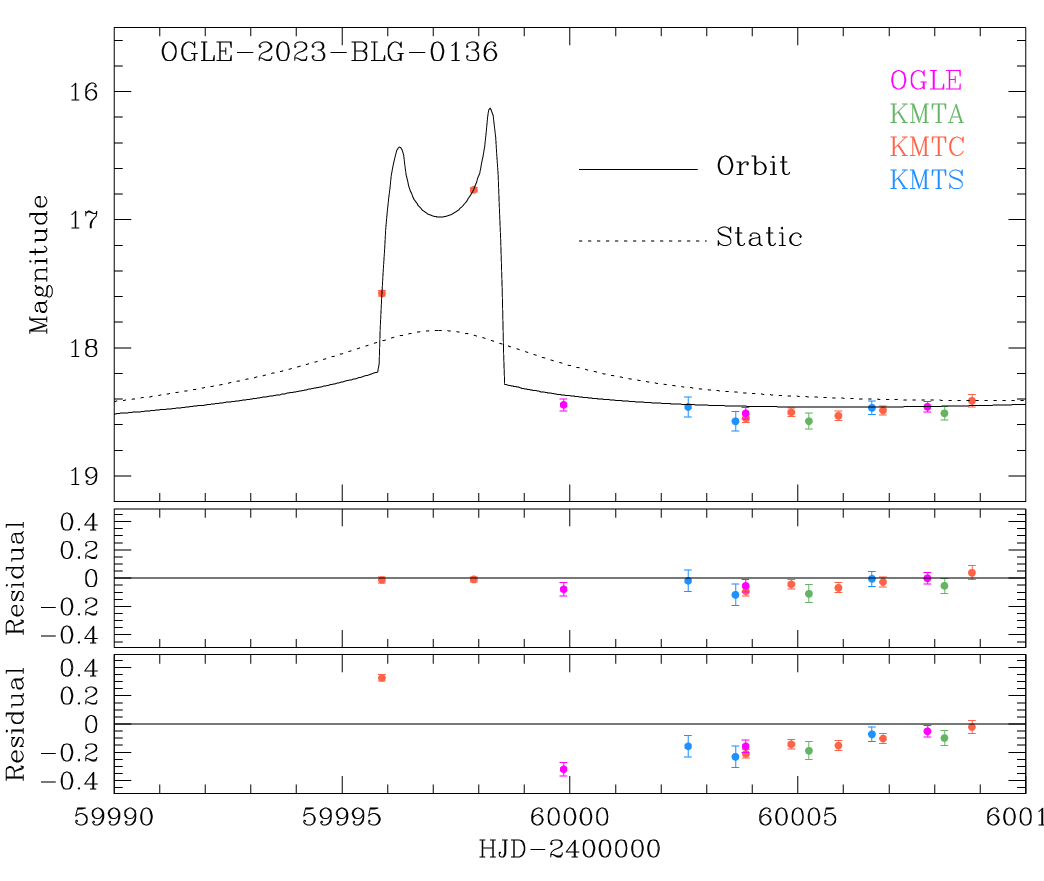}
\caption{
Enlarged view of the OGLE-2023-BLG-0136 light curve in the region around $t_1$ marked in 
Fig.~\ref{fig:eight}.
}
\label{fig:nine}
\end{figure}

We initially modeled the lensing light curve under a static 2L1S framework. The model curve of the 
static solution is depicted by a dotted curve in Figure~\ref{fig:eight}, and the lensing parameters 
of the solution are listed in Table~\ref{table:four}. Upon examination of the fit, it is observed 
that this model roughly captures the anomaly features around $t_2$ and $t_3$, yet it lacks accuracy 
in explaining the feature at $t_1$. Figure~\ref{fig:nine} offers a closer look at the model fit 
around $t_1$. While the static model does not precisely capture the first caustic-crossing anomaly, 
it does display a weak bump feature which seems to result from a caustic approach. This suggests 
that the source might pass over the caustic, potentially influenced by a slight shift in the caustic 
position due to the orbital motion of the binary lens. Considering this, we move forward with 
additional modeling, incorporating the effects of the lens orbital motion.

The solid curves in Figures~\ref{fig:eight} and \ref{fig:nine} represent the model derived 
from the orbital solution. The lensing parameters of the solution are listed in Table~\ref{table:four}. 
From the inspection of the fit, it is found that the orbital solution accurately accounts for the 
anomaly around $t_1$, improving the fit by $\Delta\chi^2=2367.4$ compared to the static solution. 
The estimated event time scale, $\te \sim 60$~days, is moderately long, and thus we further examined 
whether incorporating the microlens-parallax effect could enhance the fit. From the model derived 
with the consideration both higher-order effects, we found a marginal enhancement in the fit, 
$\Delta\chi^2=4.5$, compared to the orbital solution. This suggests that the dominant higher-order 
effect is attributed to the lens orbital motion.  Details of the lensing parameters for the parallax 
and orbit+parallax solutions are outlined in Table~\ref{table:four} and the scatter plot on the 
$\piee$--$\pien$ planet is shown in Figure~\ref{fig:three}.  The estimated binary lens parameters 
are $(s, q)\sim (0.71, 0.30)$. The normalized source radius, $\rho=(2.005\pm 0.029)\times 10^{-3}$, 
was precisely measured from the deformed light curve during the anomaly at around $t_2$. The enlarged 
view of this region is shown in the inset of Figure~\ref{fig:eight}.

\begin{figure}[t]
\includegraphics[width=\columnwidth]{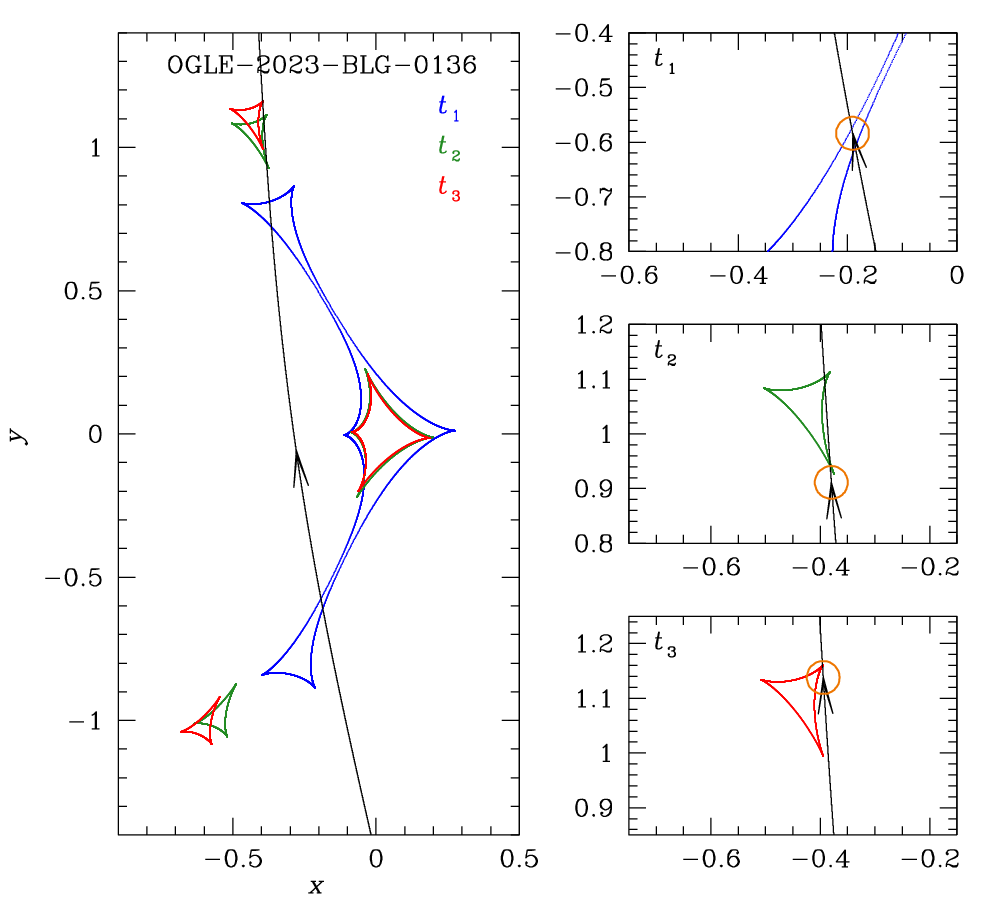}
\caption{
Configuration of the lens system for OGLE-2023-BLG-0136. 
}
\label{fig:ten}
\end{figure}

In Figure~\ref{fig:ten}, we illustrate the lens system configuration for OGLE-2023-BLG-0136. 
At the time of the first anomaly, the caustic exhibited a resonant form in which the central caustic 
and peripheral caustics were interconnected by narrow bridges. The first anomaly feature at around 
$t_1$ was generated when the source traversed the lower bridge of this resonant caustic. As the source 
proceeded, the binary separation decreased, leading to the detachment of the peripheral caustics from 
the central caustic. The second and third anomaly features arose from the successive passages of the 
source over the lower and right cusps of the upper peripheral caustic. The static model effectively 
captured the second and third anomaly features due to their close temporal proximity, which minimized 
the deformation of the caustic caused by the lens orbital motion. However, the time gap between these 
features and the first anomaly feature exceeded 100 days, resulting in substantial alterations to 
the caustic shape and position. As a result, the static model fell short in accurately representing 
the observed light curve.

It is important to note that event time scales can significantly vary between static and non-static 
models. For instance, in the case of MOA-2023-BLG-065, the time scale shifts from $\sim 30.7$~days 
for the static solution to $\sim37.8$~days for the higher-order solution. Similarly, for 
OGLE-2023-BLG-0136, the time scale changes from $\sim 66.7$~days for the static solution to 
$\sim 59.4$~days for the higher-order solution. The time scale serves as a fundamental observable 
for constraining the physical parameters of the lens. Therefore, accounting for higher-order effects 
in modeling is essential for accurately determining these parameters.

\begin{table*}[t]
\caption{Source parameters, Einstein radii, and relative proper motions \label{table:five}}
\begin{tabular}{lccccll}
\hline\hline
\multicolumn{1}{c}{Parameter}              &
\multicolumn{1}{c}{OGLE-2018-BLG-0971}     &
\multicolumn{1}{c}{MOA-2023-BLG-065}       &
\multicolumn{1}{c}{OGLE-2023-BLG-0136}     \\
\hline
 $(V-I)_{\rm S}$          &   $2.446 \pm 0.021  $   &  $1.730 \pm 0.011  $     &  $1.488 \pm 0.035  $       \\ 
 $I_{\rm S}$              &   $18.876 \pm 0.002 $   &  $20.490 \pm 0.002 $     &  $19.373 \pm 0.005 $       \\ 
 $(V-I, I)_{\rm RGC}$     &   $(2.645, 116.173) $   &  $(2.131, 15.788)  $     &  $(1.817, 15.452)  $       \\ 
 $(V-I, I)_{\rm RGC,0}$   &   $(1.060, 14.371)  $   &  $(1.060, 14.386)  $     &  $(1.060, 14.393)  $       \\ 
 $(V-I)_{\rm S,0}$        &   $0.861 \pm 0.045, $   &  $0.659 \pm 0.041, $     &  $0.731 \pm 0.053, $       \\ 
 $I_0$                    &   $17.073 \pm 0.020 $   &  $19.089 \pm 0.020 $     &  $18.315 \pm 0.021 $       \\ 
  Type                    &    K1IV                 &   G0V                    &   G4V                      \\ 
 $\theta_*$ ($\mu$as)     &   $1.433 \pm 0.119  $   &  $0.451 \pm 0.037  $     &  $0.699 \pm 0.061  $       \\ 
 $\thetae$ (mas)          &   $0.117 \pm 0.010  $   &  $0.256 \pm 0.021  $     &  $0.351 \pm 0.031  $       \\   
 $\mu$ (mas/yr)           &   $6.01 \pm 0.50    $   &  $2.42 \pm 0.20    $     &  $2.13 \pm 0.19    $       \\
\hline                                                                                                                 
\end{tabular}
\end{table*}

\section{Source stars and Einstein radii}\label{sec:four} 

In this section, we specify the source stars and determine the angular Einstein radii of the events. 
The source stars are specified based on their colors and magnitudes, accounting for corrections due 
to reddening and extinction.  In this process, we first conducted photometry of the $I$ and $V$-band 
data using the pyDIA code \citep{Albrow2017}, and then estimated the instrumental source color and 
magnitude, $(V-I, I)_{\rm S}$, by regressing the data of the individual passbands with respect to 
the model.  Calibration of the source color and magnitude follows the method outlined by 
\citet{Yoo2004}, leveraging the centroid of the red giant clump (RGC) in the color-magnitude 
diagram (CMD) for this purpose. The RGC centroid can be used for a reference because its de-reddened 
color and magnitude, represented as $(V - I, I)_{\rm RGC,0}$, are established from studies conducted 
by \citet{Bensby2013} and \citet{Nataf2013}. In the calibration process, we first positioned the source 
in the CMD constructed using the pyDIA code, measured the offsets of the source in color and magnitude, 
$\Delta (V-I, I)$, from the RGC centroid, and then estimated the de-reddened source color and magnitude 
as: 
\begin{equation}
(V - I, I)_{\rm S,0} = (V - I, I)_{\rm RGC,0} + \Delta(V - I, I). 
\label{eq3}
\end{equation}

\begin{figure}[t]
\includegraphics[width=\columnwidth]{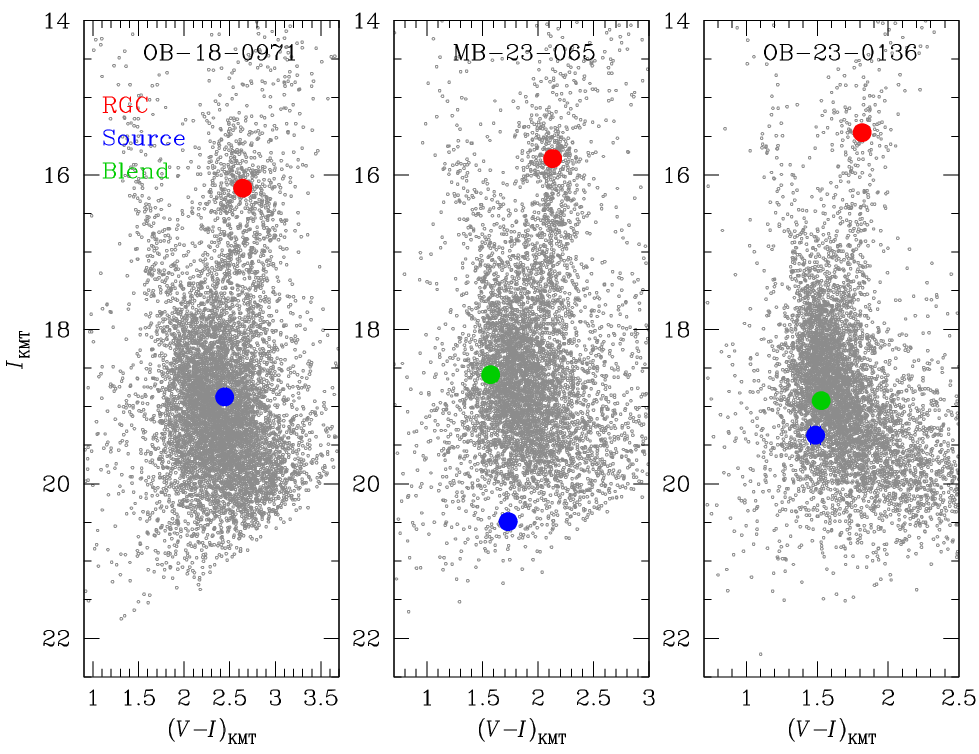}
\caption{
Positions of source stars with respect to the centroids of red giant clump (RGC) in the instrumental 
color-magnitude diagrams. For MOA-2023-BLG-065 and OGLE-2023-BLG-0136, the positions of the blends 
are additionally marked.  
}
\label{fig:eleven}
\end{figure}

In Figure~\ref{fig:eleven}, we indicate the locations of source stars for the individual events 
relative to the RGC centroids on the instrumental CMDs constructed using the KMTC data sets. The 
values estimated for $(V - I, I)_{\rm S}$, $(V - I, I)_{\rm RGC}$, $(V - I, I)_{\rm RGC,0}$, and 
$(V - I, I)_{\rm S,0}$ through the described procedure are compiled in Table~\ref{table:five}. Based 
on the derived colors and magnitudes, we determine that the source star of OGLE-2018-BLG-0971 is a 
K-type subgiant.  Additionally, the sources of MOA-2023-BLG-065 and OGLE-2023-BLG-0136 are G-type 
main-sequence stars.

The angular Einstein radius of each event was estimated from the relation
\begin{equation}
\thetae = {\theta_* \over \rho},
\label{eq4}
\end{equation}
where the angular source radius $\theta_*$ is deduced from the color and magnitude, and the normalized
source radius $\rho$ is measured from the light curve analysis. To derive the source radius, we initially
transformed the measured $V - I$ color into $V - K$ color using the \citet{Bessell1988} relation.
Subsequently, we determined $\theta_*$ using the relationship provided by \citet{Kervella2004} between
$(V - K, V)$ and $\theta_*$. The estimated angular radii of source stars for the individual lensing events
are listed in Table~\ref{table:five}, along with the corresponding angular Einstein radii calculated using 
the relationship described in Eq.~(\ref{eq3}). Also listed in the table are the relative lens-source 
proper motion estimated by the relation
\begin{equation}
\mu = {\thetae \over \te}.
\label{eq5}
\end{equation}

For MOA-2023-BLG-065 and OGLE-2023-BLG-0136, we were able to constrain the blended light.  In 
Figure~\ref{fig:eleven}, we mark the positions of blend in the CMDs. To assess the likelihood 
of the lens being the primary source of the blended flux, we measured the centroids of the source 
image during lensing magnification and at the baseline. In the case of MOA-2023-BLG-065, the measured 
astrometric offset between the source positions is $\delta\theta = 0.46 \pm 0.08$~arcsec, and this 
excludes the possibility that the blended light comes mainly from the lens. For OGLE-2023-BLG-0136, 
the significant astrometric uncertainty prevents drawing a meaningful conclusion regarding the 
origin of the blended light.

\section{Physical lens parameters}\label{sec:five} 

The physical parameters of a lens are constrained by the lensing observables of $\te$, $\thetae$, 
and $\pie$. When all these parameters are simultaneously measured, the mass and distance to the 
lens are uniquely determined by
\begin{equation}
M= {\thetae \over \kappa\pie};\qquad
\dl = { {\rm au} \over \pie\thetae + \pi_{\rm S} }.
\label{eq6}
\end{equation}
Here $\kappa = 4G/(c^2{\rm au})$ and $\pi_{\rm S} = {\rm au}/\ds$ represents the parallax of the 
source lying at a distance $\ds$ \citep{Gould2000}.  For all analyzed events, the observables of 
$\te$ and $\thetae$ were precisely measured, but the constraint on the microlens parallax was 
relatively weak because of its subtle effects.  Due to these incomplete measurements of the lensing 
observables, we determine the physical lens parameters through Bayesian analyses of the individual 
events. This approach integrates constraints from measured lensing observables with priors derived 
from the physical and dynamic distributions, as well as the mass function of lens objects within 
the Galaxy.

\begin{table*}[t]
\caption{Physical lens parameters  \label{table:six}}
\begin{tabular}{lccccll}
\hline\hline
\multicolumn{1}{c}{Parameter}              &
\multicolumn{1}{c}{OGLE-2018-BLG-0971}     &
\multicolumn{1}{c}{MOA-2023-BLG-065}       &
\multicolumn{1}{c}{OGLE-2023-BLG-0136}     \\
\hline
 $M_1$ ($M_\odot$)    &    $0.215^{+0.231}_{-0.126}$   &  $0.30^{+0.27}_{-0.15}$   &   $0.88^{+0.23}_{-0.18}   $    \\ [0.8ex]
 $M_2$ ($M_\odot$)    &    $0.186^{+0.200}_{-0.109}$   &  $0.28^{+0.25}_{-0.14}$   &   $0.259^{+0.068}_{-0.052}$    \\ [0.8ex]
 $\dl$ (kpc)          &    $7.50^{+0.98}_{-1.01}   $   &  $7.62^{+0.89}_{-0.84}$   &   $7.52^{+0.92}_{-1.13}   $    \\ [0.8ex]
 $a_\perp$ (au)       &    $1.39^{+0.18}_{-0.19}   $   &  $3.10^{+0.36}_{-0.34}$   &   $1.92^{+0.23}_{-0.29}   $    \\ [0.8ex]
 $p_{\rm disk}$       &    $26\%                   $   &  $9\%                 $   &   $21\%                   $    \\ [0.8ex]
 $p_{\rm bulge}$      &    $74\%                   $   &  $91\%                $   &   $79\%                   $    \\ [0.8ex]
\hline                                                                                                                 
\end{tabular}
\end{table*}

The Bayesian analysis began by generating a large number of synthetic events via Monte Carlo
simulation. Within this simulation, the physical parameters of the lens mass were deduced from 
a model mass function, while the distances to the lens and source, along with their relative 
proper motion, were derived from a Galaxy model. Our approach incorporated the mass function 
model suggested by \citet{Jung2018} and utilized the Galaxy model introduced by \citet{Jung2021}.  
For each synthetic event defined by physical parameters $(M_i, D_{{\rm L},i}, D_{{\rm S},i}, 
\mu_i)$, we calculated the the values of the corresponding lensing observables using the relations 
\begin{equation}
t_{{\rm E},i} = {\theta_{{\rm E},i} \over \mu_i};\qquad
\theta_{{\rm E},i} = \sqrt{\kappa M_i \pi_{{\rm rel},i}}
\label{eq7}
\end{equation}
for the event time scale and Einstein radius, respectively, and using the relation in Eq.~(\ref{eq1})
for the microlens parallax.  Then, the posteriors of $M$ and $\dl$ were obtained by assigning a 
weight to each event of $w_i = \exp (-\chi_i^2/2)$, where $\chi_i^2$ value is computed by 
\begin{equation}
\chi_i^2 = 
{ \Delta t_{{\rm E},i}^2\over \sigma^2(\te)} + 
{ \Delta\theta_{{\rm E},i}^2\over \sigma^2(\thetae)} +
\sum_{j=1}^2 \sum_{k=1}^2 b_{j,k}
(\pi_{{\rm E},j,i} - \pi_{{\rm E},i})
(\pi_{{\rm E},k,i} - \pi_{{\rm E},i}).
\label{eq8}
\end{equation}
Here, $\Delta t_{{\rm E},i}=t_{{\rm E},i} - \te$, $\Delta\theta_{{\rm E},i}=\theta_{{\rm E},i} - 
\thetae$, $(\te, \thetae)$ stand for the measured values of the observables, $[\sigma(\te), 
\sigma(\thetae)]$ indicate their corresponding uncertainties, and $b_{j,k}$ represents the 
inverse covariance matrix of the microlens-parallax vector $\pivec_{\rm E}$, $(\pi_{{\rm E},1}, 
\pi_{{\rm E},2})_i = (\pien, \piee)_i$ are the parallax parameters of each simulated event, and 
$(\pien, \piee)$ represent the parallax parameters measured from modeling.  \citet{Han2016b} 
showed that parallax measurements can be important even when there is little $\chi^2$ improvement, 
because they can constrain $\pie$ to be small.  We, therefore,  consider the constraints given 
by the measured parallax parameters in the Bayesian analyses.

\begin{figure}[t]
\includegraphics[width=\columnwidth]{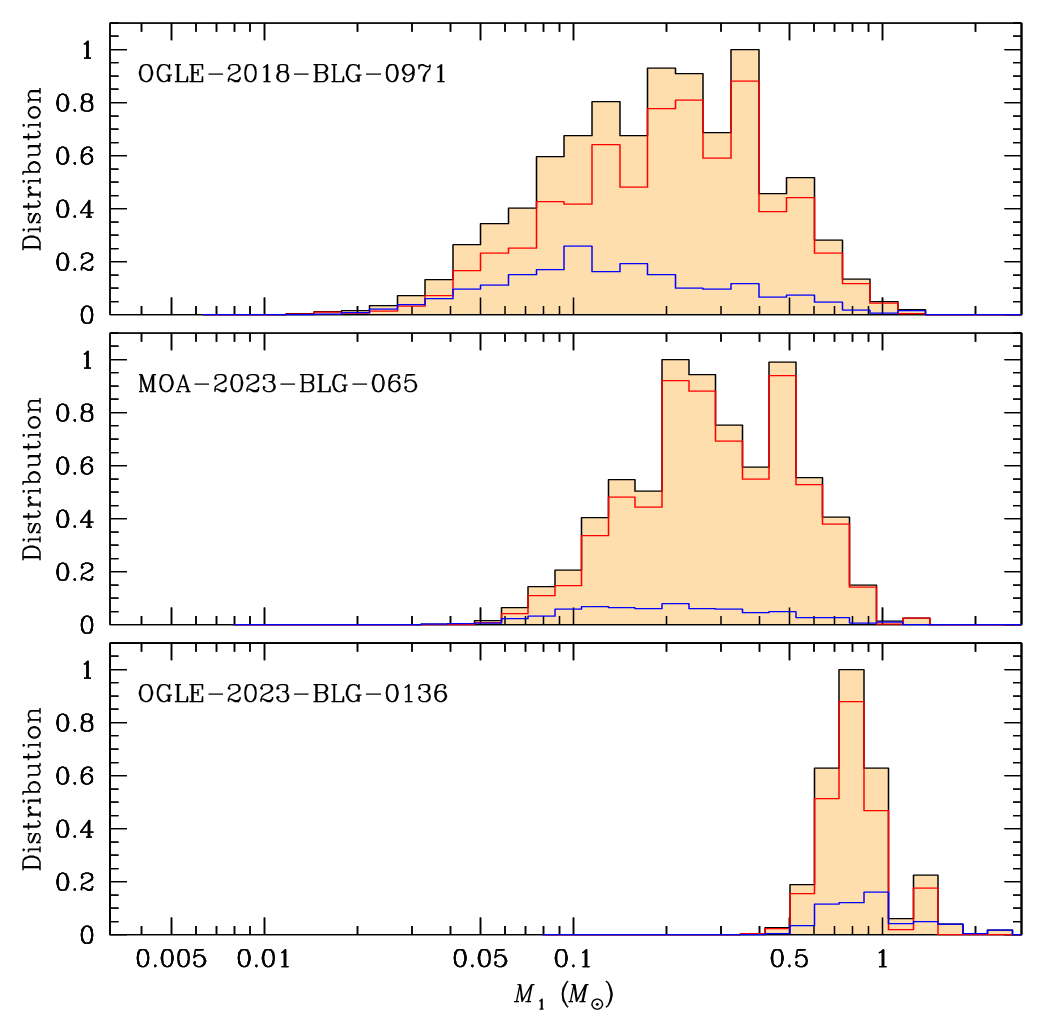}
\caption{
Bayesian posterior distributions of the primary lens mass ($M_1$) for the lensing events. In 
each panel, the event rate contributions from the disk and bulge lens populations are depicted 
by the blue and red curves, respectively, while black curve represents the sum of the contributions 
from two lens populations.
}
\label{fig:twelve}
\end{figure}

\begin{figure}[t]
\includegraphics[width=\columnwidth]{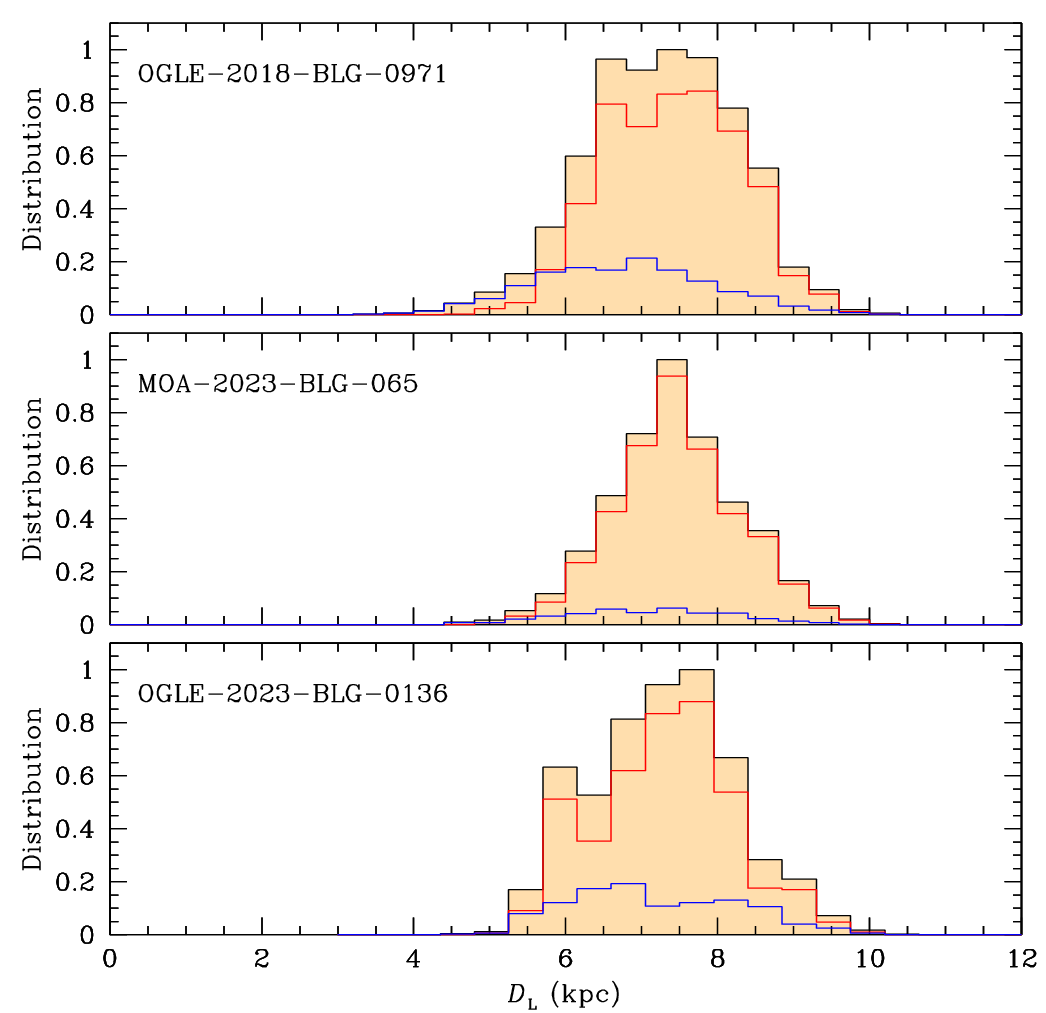}
\caption{
Bayesian posterior distributions of the distances to the lens ($\dl$). The notations correspond to
those used in Fig.~\ref{fig:twelve}.
}
\label{fig:thirteen}
\end{figure}

In Figures~\ref{fig:twelve} and \ref{fig:thirteen}, we present the Bayesian posteriors of 
the primary lens masses and distances to the lens systems.  In Table~\ref{table:six}, we 
summarize the estimated physical parameters for the individual lensing events.  Among the 
parameters, $M_1$ and $M_2$ denote the masses of the primary and companion of the lens, and 
$a_\perp$ denotes the projected physical separation between $M_1$ and $N_2$.  We present the 
median value derived from the Bayesian posterior distribution as a representative value for 
each physical parameter, with uncertainties estimated within the 16\% to 84\% range of the 
distribution.  Also listed in the table are the relative probabilities for the lens being in 
the Galactic disk, $p_{\rm disk}$, and in the bulge, $p_{\rm bulge}$.  According to the 
estimated masses, the lenses of the events OGLE-2018-BLG-0971 and MOA-2023-BLG-065 are binaries 
composed of M dwarfs.  On the other hand, the lens of OGLE-2023-BLG-0136 is likely to be a binary 
composed of an early K-dwarf primary and a late M-dwarf companion.  Across all the analyzed events, 
$p_{\rm bulge}$ is substantially higher than $p_{\rm disk}$, suggesting a higher likelihood of 
the all lenses being located in the bulge rather than the disk.  The probabilities $p_{\rm disk}$ 
and $p_{\rm bulge}$ were found by analyzing the proportion of artificial lensing events where the 
lenses came from either the disk or bulge distributions within the Galaxy model utilized in the 
Monte Carlo simulation process.

\section{Summary}\label{sec:six}

We have analyzed microlensing data collected from high-cadence surveys to reevaluate lensing
events that lacked proposed interpretations for their intricate anomaly features. Through detailed
reanalyses considering higher-order effects, we identified that accounting for the orbital motions of
lenses was vital in accurately explaining the anomaly features observed in the lensing light curves
of OGLE-2018-BLG-0971, MOA-2023-BLG-065, and OGLE-2023-BLG-0136.

By conducting Bayesian analyses based on the lensing parameters from the newly found solutions
and together with the constraints derived from lensing observables, we estimated masses and
distances to the lenses.  These analyses revealed that the lenses for events 
OGLE-2018-BLG-0971 and MOA-2023-BLG-065 are binary systems consisting of M dwarfs.  Additionally, 
for OGLE-2023-BLG-0136, the lens is likely to be a binary system comprising an early K-dwarf primary 
and a late M-dwarf companion.  Notably, across all observed lensing events, the likelihood of the 
lens being in the bulge significantly outweighs its likelihood of being in the disk.

\begin{acknowledgements}
Work by C.H. was supported by the grants of National Research Foundation of Korea (2019R1A2C2085965). 
J.C.Y. and I.-G.S. acknowledge support from U.S. NSF Grant No. AST-2108414. 
Y.S. acknowledges support from BSF Grant No. 2020740.
This research has made use of the KMTNet system operated by the Korea Astronomy and Space
Science Institute (KASI) at three host sites of CTIO in Chile, SAAO in South Africa, and SSO in
Australia. Data transfer from the host site to KASI was supported by the Korea Research
Environment Open NETwork (KREONET). This research was supported by KASI under the R\&D
program (project No. 2023-1-832-03) supervised by the Ministry of Science and ICT.
W.Z. and H.Y. acknowledge support by the National Natural Science Foundation of China (Grant
No. 12133005).
W. Zang acknowledges the support from the Harvard-Smithsonian Center for Astrophysics through
the CfA Fellowship. 
The MOA project is supported by JSPS KAKENHI Grant Number JP24253004, JP26247023, JP16H06287 
and JP22H00153.
\end{acknowledgements}

\end{document}